      [1999/12/01 v1.4c Il Nuovo Cimento]
\documentclass{cimento}

\usepackage{graphicx}

\title{Nucleation and evolution of false vacuum bubbles in scalar-tensor gravity\thanks{A proceeding for the 12th Italian-Korean Symposium on Relativistic Astrophysics. Talk on the 7th of July, 2011, Pescara, Italy.}}
\author{Bum-Hoon Lee\from{ins:x}\thanks{\texttt{bhl@sogang.ac.kr}} \atque Dong-han Yeom\from{ins:x}\from{ins:y}\thanks{\texttt{innocent.yeom@gmail.com}}}
\instlist{\inst{ins:x}Center for Quantum Spacetime, Sogang University, Seoul 121-742, Republic of Korea
\inst{ins:y}Research Institute for Basic Science, Sogang University, Seoul 121-742, Republic of Korea}
\PACSes{PACS: 98.80.Cq, 04.70.Dy, 04.62.+v, 11.27.+d}
\begin{document}

\maketitle

\begin{abstract}
In this presentation, we discuss the nucleation and subsequent evolution of false vacuum bubbles in the scalar-tensor gravity. First, we transform the scalar-tensor type theory of gravity to the standard Brans-Dicke type. Second, we transform the Brans-Dicke type theory from the Jordan frame to the Einstein frame. For a certain potential, a true vacuum bubble in the Einstein frame can be transformed to a false vacuum bubble in the Jordan frame by a conformal transformation. Thus, in the Jordan frame, the nucleation of a false vacuum bubble can be possible and its subsequent evolution can be described with the help of thin-wall approximation. False vacuum bubbles have physical importance: a set of false vacuum bubbles might generate a negative energy bath and it has further theoretical implications.
\end{abstract}

\section{Introduction}

Recently, the application of modified gravity for general relativity has gained vivid interest. People are interested in the following questions: What is the motivation behind such a modified gravity theory? What are the observational consequences of the theory for black holes, cosmology or other areas? What is the theoretical implication of the modified gravity theory?

In this context, the scalar-tensor models are the most promising candidate, since they generally have quite clear theoretical motivations. In this paper, we will discuss the false vacuum bubbles in the scalar-tensor gravity. This issue is related to the following interesting and important problems:
\begin{itemize}
\item \textsf{Cosmology:} Is the bubble universe via false vacuum bubbles viable with the scalar-tensor gravity?
\item \textsf{General relativity:} Is the Jordan frame equivalent to the Einstein frame, in the presence of false vacuum bubbles?
\item \textsf{String theory:} If a false vacuum bubble is allowed, what is its implication to holography?
\item \textsf{Quantum gravity:} If a false vacuum bubble or a bubble universe exists, does it cause the information loss problem?
\end{itemize}
Of course, these are too difficult questions, and hence it is difficult to give an ultimate answer to some of the questions; however, in this paper, we want to assert the fact that \textit{a false vacuum bubble in the scalar-tensor gravity is a \textbf{good toy model} to study these problems}.

\subsection{Scalar-tensor models}
The prototype of a scalar-tensor model is the Brans-Dicke theory \cite{brdi01} given by the action:
\begin{eqnarray}
S = \frac{1}{16\pi} \int \sqrt{-g} d^4 x \left( \Phi R - \omega g^{\alpha\beta} \frac{\nabla_{\alpha}\Phi
\nabla_{\beta}\Phi}{\Phi} + V(\Phi) \right).
\end{eqnarray}
Here, $\omega$ is a dimensionless coupling parameter, and the Einstein gravity is restored back in the $\omega \rightarrow \infty$ limit. $V(\Phi)$ is the potential for the Brans-Dicke field $\Phi$.

Matching current observations with the Brans-Dicke theory without a potential requires $\omega$ to be greater than $\sim 40,000$ \cite{Ber}. However, the study for small $\omega$ parameters have theoretical interests for three distinct reasons. $(1)$ If a potential is involved with the Brans-Dicke field, then the dynamics of the Brans-Dicke field is restricted and hence one can obtain a viable model even though $\omega$ is smaller than $40,000$. $(2)$ Via the \textit{chameleon mechanism}, an effective potential can be derived when there is a local matter density around the solar system \cite{Khoury:2003aq}; this can in turn make $\omega$ apparently look large for solar system experiments. $(3)$ Some string-inspired models allow various values for $\omega$ and hence the study for small $\omega$ universe has got some theoretical importance for understanding the nature of string theory \cite{Fujii:2003pa}. For example, dilaton gravity can be restored from string theory as the $\omega=-1$ limit. In some cases, higher order corrections in string theory can make the dimensionless coupling to be field dependent: $\omega(\Phi)$. The Randall-Sundrum model \cite{Randall:1999ee} allows the weak field limit of gravity on the brane to be $\omega = (3/2) [\exp{(\pm s/l)} - 1]$ \cite{Garriga:1999yh}. Moreover, apart from string theory, higher order curvature corrections can be described by scalar-tensor models; for example, in $f(R)$ gravity, it is equivalent to the $\omega=0$ limit with a suitable potential \cite{Sotiriou:2008rp}.

\subsection{Conformal transformations}

Almost all scalar-tensor models can be transformed into the prototype Brans-Dicke theory via field redefinitions. We can further transform the model to the Einstein theory if $\omega > -3/2$. This is known to be a conformal transformation \cite{Fujii:2003pa}:
\begin{eqnarray}
\bar{g}_{\mu\nu} = \Phi g_{\mu\nu}, \;\;\; \bar{\phi} = \sqrt{\frac{2\omega + 3}{16 \pi}} \ln \Phi, \;\;\; \bar{U}(\bar{\phi}) = \Phi^{-2} V(\Phi),
\end{eqnarray}
where $\bar{g}_{\mu\nu}$ is the metric in the Einstein frame and $\bar{\phi}$ is the minimally coupled scalar field in the Einstein frame with the potential $\bar{U}(\bar{\phi})$.

\section{Nucleation of false vacuum bubbles}

In order to describe the nucleation process, there is a well known procedure \cite{CDL} which uses the $O(4)$ symmetric metric in the Euclidean signature $ds^{2} = d\eta^{2} + \rho^{2}(\eta)d\Omega^{2}_{3}$, and then obtains the bounce solution for the field $\Phi$ and the scale factor $\rho$. Unfortunately, it is well-known that a small false vacuum bubble\footnote{That is, the size of the bubble is smaller than the size of the cosmological horizon of the background.} in the Einstein gravity is not allowed. Then, one might guess that scalar-tensor gravity will not allow any small false vacuum bubble, since scalar-tensor models can be transformed to the Einstein frame. Of course, in terms of considerations of the Einstein frame, small false vacuum bubbles are impossible. However, \textit{in the Jordan frame, this may not be true.} That is, it is possible that \textit{a true vacuum bubble in the Einstein frame can be a false vacuum bubble in the Jordan frame}.

If this has to happen, we have to require some simple conditions on the potential \cite{Kim:2010yr}: $V(\Phi_{\mathrm{t}}) < V(\Phi_{\mathrm{f}})$ and $\bar{U}(\bar{\phi}_{\mathrm{t}}) > \bar{U}(\bar{\phi}_{\mathrm{f}})$, where subscripts $\mathrm{t}$ and $\mathrm{f}$ respectively denote the true and the false vacuum in the Jordan frame. For convenience, we define the effective force function as $F(\Phi) = \Phi V'(\Phi) - 2V(\Phi)$, choosing $\Phi_{\mathrm{t}}=1$ and $V(\Phi_{\mathrm{t}})=V_{0}$, and we represent the conditions to be
\begin{eqnarray}
V(\Phi_{\mathrm{f}})-V(\Phi_{\mathrm{t}}) = \Phi_{\mathrm{f}}^{2} \left( \int_{1}^{\Phi_{\mathrm{f}}} \frac{F(\bar{\Phi})}{\bar{\Phi}^{3}} d \bar{\Phi} + V_{0} \right) - V_{0} > 0
\end{eqnarray}
and
\begin{eqnarray}
\bar{U}(\bar{\phi}_{\mathrm{f}})-\bar{U}(\bar{\phi}_{\mathrm{t}}) = \bar{U}(\Phi_{\mathrm{f}})-\bar{U}(\Phi_{\mathrm{t}}) = \int_{1}^{\Phi_{\mathrm{f}}} \frac{F(\bar{\Phi})}{\bar{\Phi}^{3}} d \bar{\Phi} \equiv \Delta E < 0.
\end{eqnarray}
Therefore, we require $V_{0} > \Phi^{2}_{\mathrm{f}} |\Delta E|/ (\Phi^{2}_{\mathrm{f}}-1)$ and we conclude that such false vacuum bubbles can form only in a de Sitter background with $V_{0}>0$ and $\Phi_{\mathrm{f}} > 1$. An explicit example of such a potential and a false vacuum bubble solution was reported in \cite{Kim:2010yr}.

For explicit calculations, we use the thin-wall approximation
\begin{eqnarray}
\dot{\Phi}\frac{\dot{\rho}}{\rho} \ll 1.
\end{eqnarray}
Then, we can approximate the Euclidean action and obtain
\begin{eqnarray}
B = B_{\mathrm{outside}} + B_{\mathrm{wall}} + B_{\mathrm{inside}},
\end{eqnarray}
where $B_{...} = B(...|\mathrm{bounce}) - B(...|\mathrm{background})$. The size of the thin-wall $\bar{\rho}$ should then satisfy the following equation of motion
\begin{eqnarray}
\frac{\partial B}{\partial \bar{\rho}} = 0 = \frac{3 \pi}{2} \bar{\rho} \left( 4 \pi \bar{\rho} \sigma_{0}- \Phi_{\mathrm{f}} \sqrt{f_{\mathrm{f}}} + \Phi_{\mathrm{t}} \sqrt{f_{\mathrm{t}}} \right),
\end{eqnarray}
where $\sigma_{0}$ is a constant tension parameter and $f_{\mathrm{t,f}} = 1- (V(\Phi_{\mathrm{t,f}})/6\Phi_{\mathrm{t,f}})\bar{\rho}^{2}$ \cite{Kim:2010yr}. Note that the Junction condition for the Lorentzian signature can be represented by \cite{Lee:2010yd}
\begin{equation} \label{eq:junction}
\epsilon_{\mathrm{f}} \Phi_{\mathrm{f}}\sqrt{\dot{\bar{\rho}}^2 + f_{\mathrm{f}}} - \epsilon_{\mathrm{t}} \Phi_{\mathrm{t}} \sqrt{\dot{\bar{\rho}}^2 + f_{\mathrm{t}}} = 4\pi \bar{\rho} \sigma_{0}.
\end{equation}
Therefore, we can smoothly connect from the bounce solution to the subsequent evolution at the $t=0$ surface, by choosing $\epsilon_{\mathrm{t,f}}=+1$.

\section{Subsequent dynamics of false vacuum bubbles}

According to the discussion of the previous section, we can sure that the junction equation (Equation~(\ref{eq:junction}), \cite{Lee:2010yd}) is the thin-wall limit of the Euclidean thick-wall solution. Now, we will assume $\epsilon_{\mathrm{t,f}}=+1$ as we already observed from the Euclidean solution, although it is not necessarily true in general. Then we can rewrite the equation as
\begin{equation}
\sqrt{\dot{\bar{\rho}}^2 + f_{\mathrm{f}}} - \sqrt{\dot{\bar{\rho}}^2 + f_{\mathrm{t}}} = 4\pi \bar{\rho} \left(\sigma_{0} + \frac{(1-\Phi_{\mathrm{f}})}{4\pi \bar{\rho}}\sqrt{\dot{\bar{\rho}}^2 + f_{\mathrm{f}}} \right),
\end{equation}
where we have fixed $\Phi_{\mathrm{t}}=1$ without loss of generality. Then we can rewrite the right hand side as $4\pi \bar{\rho} \sigma_{\mathrm{eff}}(\bar{\rho})$, where $\sigma_{\mathrm{eff}}(\bar{\rho})$ is the effective tension. In this rewritten form, we can calculate the Lorentzian dynamics of the thin-wall, as we already calculated in the Einstein case. Since $\Phi_{\mathrm{f}}>1$, the effective tension can be negative, and it is not difficult to check that the effective tension indeed becomes negative \cite{Lee:2010yd}\cite{lll2006}.

\begin{figure}
\begin{center}
\includegraphics[scale=0.55]{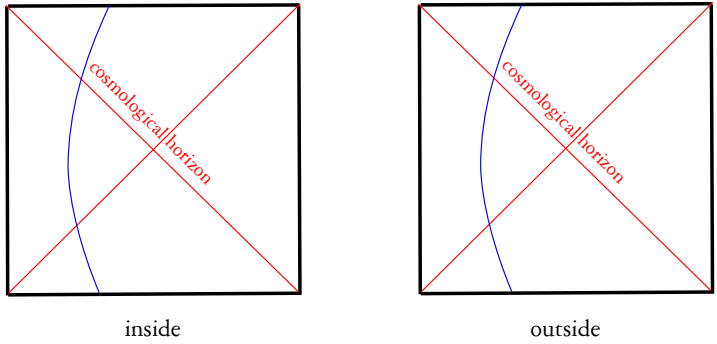}
\caption{\label{fig:thinshell}Subsequent dynamics of the bubble is depicted above. In the Jordan frame, the size of the cosmological horizon of the inside is smaller than that of the outside; whereas in the Einstein frame, the size of the cosmological horizon of the inside is larger than that of the outside.}
\end{center}
\end{figure}

\begin{figure}
\begin{center}
\includegraphics[scale=0.55]{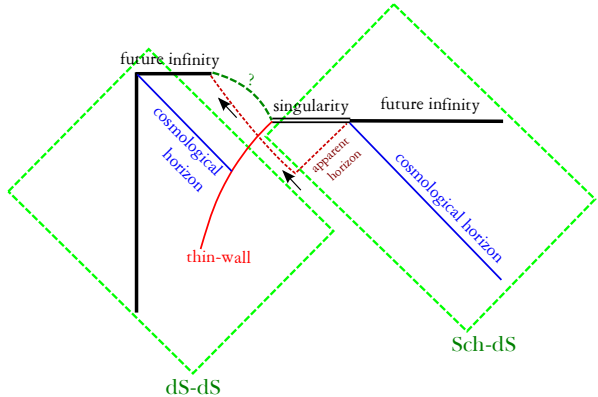}
\caption{\label{fig:diagram}Formation of a black hole in the Jordan frame in the presence of a false vacuum bubble. This is obtained by pasting two solutions (two green boxes): a false vacuum bubble (dS-dS) and a de Sitter black hole (Sch-dS).} 
\end{center}
\end{figure}

The fact that the tension becomes negative is important to decide the causal structure \cite{Blau:1986cw}. A true vacuum bubble wall can expand to the asymptotic observer, but the same is impossible for a false vacuum bubble if the tension is positive \cite{Blau:1986cw}\cite{Lee:2010yd}. Therefore, we can interpret this in two ways: (1) Our solution is a false vacuum bubble in the Jordan frame, and it expands over the asymptotic observer via negative tension and (2) our solution is a true vacuum bubble in the Einstein frame, and it expands over the asymptotic observer via positive tension (the latter case is trivial). These two interpretations are of course consistent. Figure~\ref{fig:thinshell} explains this \cite{Kim:2010yr}. In the Jordan frame, the size of the cosmological horizon of the inside is smaller than that of the outside; whereas in the Einstein frame, the size of the cosmological horizon of the inside is larger than that of the outside.


We can think of a limit such that the cosmological horizon of a false vacuum bubble in the Jordan frame $l_{\mathrm{f}}$ is much smaller than that of the true vacuum $l_{\mathrm{t}}$. This might happen for the $V_{0} < |\Delta E| \ll 1$ and $\Phi_{\mathrm{f}} \gg 1$ limit and there is no reason to prohibit this combination to happen. Then, the formation of a black hole with mass $M$ in the Jordan frame can be allowed: $l_{\mathrm{f}} < 2 M < l_{\mathrm{t}}$. Dynamical formation of such a black hole is obtained to paste the solution from `a false vacuum bubble in a true vacuum (dS-dS)' to `a black hole in a de Sitter space (Sch-dS)'(Figure~\ref{fig:diagram}). In the thin-wall limit, during the formation of a black hole, the corresponding solution in the Einstein frame is less clear; and hence, this should be discussed using the complete thick-wall solutions.


\section{Lessons from double-null formalism}

To describe the dynamics of the shell beyond the thin-wall approximation, we may need to employ numerical calculations. The authors and colleagues used the double-null formalism for this purpose \cite{Hong:2008mw}. The double-null formalism is based on the double-null coordinates: $ds^{2} = -\alpha^{2}(u,v)dudv + r^{2}(u,v)d\Omega^{2}$. Then, we formulate the Einstein equations and matter field equations in this coordinates and solve them numerically\footnote{In this short proceeding, we cannot discuss the detailed method of the simulations and hence interested readers can look into \cite{Hong:2008mw} for further details.}.

What we can infer from this numerical calculations for bubble dynamics are as follows:
\begin{enumerate}
\item A collapsing false vacuum bubble can emit negative energy along the out-going direction via semi-classical effects, such that it can be a source of negative energy fluxes \cite{Yeom:2009mn}\cite{Hwang:2010gc}.
\item If there is a violation of the null energy condition or if the negative energy is concentrated on the wall, then a false vacuum bubble can expand over the asymptotic region \cite{Hansen:2009kn}.
\end{enumerate}
The former result is not a strange one, and we can explicitly show by numerical calculations. The latter is a trivial result, and we can explicitly demonstrate via numerical computations.

\begin{figure}
\begin{center}
\includegraphics[scale=0.3]{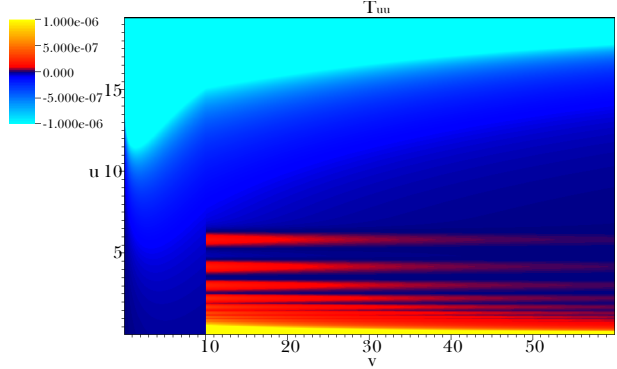}
\caption{\label{fig:various}A collapsing false vacuum bubble emits negative energy such that the out-going energy density $T_{uu}$ becomes negative.}
\end{center}
\end{figure}

In the previous work, we approximated the semi-classical energy-momentum tensor using the $S$-wave approximation: we use two-dimensional results for $\langle T_{\mu\nu}\rangle$ and divide by $4\pi r^{2}$. There is indeed some well-developed justification in support of the use of the $S$-wave approximation for the black hole cases \footnote{Refs. \cite{Hong:2008mw}\cite{Hwang:2010im} and references therein include qualitative justifications for this approximation.}. The emission of the negative energy flux is not entirely a new phenomenon, since we can estimate $\langle T_{\mu\nu}\rangle$ of the two-dimensional de Sitter space which is $\sim - \Lambda$, where $\Lambda$ is the cosmological constant of the de Sitter space \cite{Yeom:2009mn}\cite{Davies:1976ei}. Effectively, the outside of a cosmological horizon of a de Sitter space is filled with a negative energy thermal particles, and therefore, if there is a false vacuum bubble, it will effectively emit negative energy particles, since the temperature of the false vacuum is higher than the true vacuum.

Figure~\ref{fig:various} shows that a collapsing false vacuum bubble can emit negative energy flux along the out-going direction (Figure~12 in \cite{Hwang:2010gc}). $v < 10$ is the false vacuum region and $v > 10$ is the true vacuum region. The wall between two regions is located around $v \approx 10$. Then the shell is formed towards an in-going null direction and will collapse toward the center. The blue region is where the null energy condition is violated, i.e., $T_{uu}$ (the out-going part of the energy-momentum tensor) becomes negative, and hence the negative energy is emitted along the out-going null direction. Asymptotic region of $T_{uu}$ is yellow, i.e., there are some part with positive out-going energy; however, we could see that the out-going energy can be controlled by initial conditions \cite{Hwang:2010gc}.

Now we want to make some comments on the simulation. In the previous calculation, we assumed a bubble in an almost flat background. Therefore, the choice for the asymptotic mass was always approximately zero. However, in many situations, the asymptotic mass can be different; then, the tendency for the emission of negative energy can be changed. However, it is a definite fact that there is a solution of Einstein gravity that allows a false vacuum bubble to emit negative energy, at least in the $S$-wave approximation level.

\begin{figure}
\begin{center}
\includegraphics[scale=0.3]{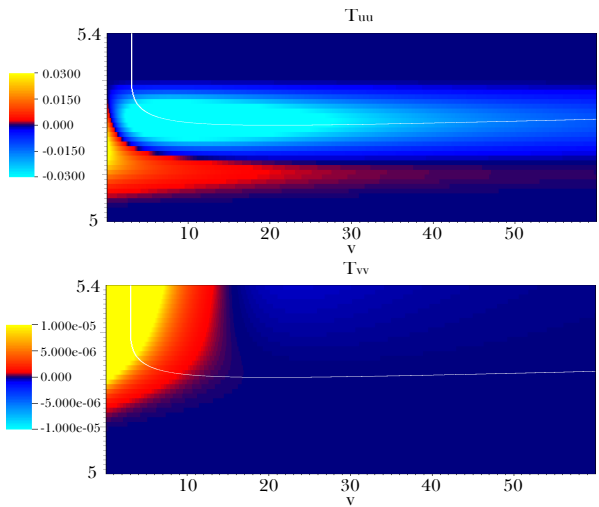}
\caption{\label{fig:sol3_shell}If the null-energy condition $T_{uu}$ is violated on the wall, then inflation (the existence of an anti-trapping horizon (white curve)) is possible.}
\end{center}
\end{figure}

Figure~\ref{fig:sol3_shell} is an example of the expanding and inflating wall (Figure~19 in \cite{Hansen:2009kn}). In this case, $v < 5$ is the true vacuum region and $v > 5.4$ is the false vacuum region; hence, the wall is formed towards an out-going direction. This figure explicitly shows, when the null-energy condition $T_{uu}$ on the wall is violated or the negative energy is concentrated on the wall, the existence of inflation, i.e., the existence of an anti-trapping horizon (white curve) is possible. Perhaps, collapsing false vacuum bubbles can be the origin of the negative energy; and \textit{collapsing false vacuum bubbles may induce an expanding false vacuum bubble} \cite{Yeom:2009mn}\cite{Hwang:2010gc}.

\section{Conclusion}

In this paper, we discussed the nucleation of a false vacuum bubble in the scalar-tensor gravity, especially in the Jordan frame. The existence of such a false vacuum bubble is not contradictory, since we can match the solution with a true vacuum bubble in the Einstein frame \cite{Kim:2010yr}. We can continuously connect the Euclidean solution \cite{Kim:2010yr} to the Lorentzian solution \cite{Lee:2010yd}, and perhaps, the solution can allow a formation of the bubble universe that is separated from the asymptotic region (Figures~\ref{fig:thinshell} and \ref{fig:diagram}).

There are some supplementary evidences for the existence of a bubble universe. The nucleated false vacuum bubbles can be used as sources of negative energy fluxes (Figure~\ref{fig:various}, \cite{Yeom:2009mn}\cite{Hwang:2010gc}). If the negative energy is concentrated on a wall of a false vacuum bubble, then it will allow the formation of a bubble universe (Figure~\ref{fig:sol3_shell}, \cite{Hansen:2009kn}), even though the bubble initially satisfies the null energy condition.

As answers for the questions discussed in the introduction, we have the following comments:
\begin{itemize}
\item The existence of a bubble universe separated from the asymptotic region seems to be possible. At least, in the Jordan frame, we have a definite solution $l_{\mathrm{f}}<l_{\mathrm{t}}$ (Figure~\ref{fig:diagram}). If one observer sees the bubble at $r$ with the condition $l_{\mathrm{f}}< \bar{\rho} < r <l_{\mathrm{t}}$, then we can choose a space-like hypersurface such that the observer is inside the cosmological horizon and there is the second asymptotic region inside the observer along the space-like hypersurface.
\item The next natural question to address is the meaning associated with the information loss problem. It is not sufficient to say something regarding the information loss problem using only the false vacuum bubble; for example, the entropy of the inside region is smaller than that of the outside (note that, the entropy should be calculated in the Einstein frame), and hence it is not entirely inconsistent with the unitarity. If we clearly separate the space-time by a black hole (Figure~\ref{fig:diagram}), then the loss of information will be clearer. However, still there remains a choice of black hole complementarity or holography \cite{Susskind:1993if}\cite{Yeom:2008qw}. Also, one can ask whether we can apply the arguments of the fuzzball conjecture or not \cite{Mathur:2005zp} in this context.
\item If we include the semi-classical effects, the outside of a false vacuum bubble will emit negative energy along the out-going direction. However, outside of a true vacuum bubble will not emit negative energy, since the background temperature of the false vacuum region is hotter than the true vacuum region. Therefore, a false vacuum bubble and a true vacuum bubble can be distinguished by semi-classical effects. Then, although our false vacuum bubble solution in the Jordan frame is classically equivalent to the true vacuum bubble in the Einstein frame, they can be distinguished by including semi-classical effects.
\item If there is an emission of negative energy, it can be concentrated on a wall of a false vacuum bubble, and it can in turn induce an expanding and inflating false vacuum bubble. Then, it can produce a supplementary evidence for the existence of a bubble universe that is separated from the asymptotic region.
\end{itemize}

Thus, the nucleation and evolution of false vacuum bubbles in the scalar-tensor gravity will give us a good toy model to investigate for the existence of a bubble universe and the information loss problem of black holes. The existence of a bubble universe becomes clearer, but the implication on the information loss problem and holography in string theory is not so clear. Emission of negative energy from a false vacuum bubble is related to the problem whether the Jordan frame is equivalent to the Einstein frame or not. It seems that they are not equivalent at the quantum level; this fact should be studied in detail and should be supported with more concrete calculations, beyond the $S$-wave approximation.

\acknowledgments
The authors would like to thank Hongsu Kim, Wonwoo Lee, Jakob Hansen, Dong-il Hwang, and Young Jae Lee for discussions and encouragement. The authors especially thank Remo Ruffini and Hyung Won Lee for the hospitality we received during the 12th Italian-Korean Symposium on Relativistic Astrophysics. DY thanks Raju Roychowdhury and Chaitali Roychowdhury for the careful English revision. This work is supported by the National Research Foundation of Korea (NRF) grant funded by the Korea government(MEST) through the Center for Quantum Spacetime (CQUeST) of Sogang University with grant number 2005-0049409.

\end{document}